\documentclass[twocolumn,showpacs,prb,amsmath,amssymb]{revtex4}

\usepackage{graphicx}\usepackage{dcolumn}\usepackage{bm}

\begin{document}

\title{Ferrimagnetic mixed-spin ladders in weak and strong coupling limits}

\author{D.N.Aristov}
\altaffiliation[On leave from ]
{Petersburg Nuclear Physics Institute, Gatchina  188300, Russia.}
\affiliation{
Max-Planck-Institute   FKF, Heisenbergstra\ss e 1, 70569 Stuttgart,
Germany}\author{M.N. Kiselev}
\affiliation{ Institut f\"ur Theoretische Physik und Astrophysik,
Universit\"at W\"urzburg,
 D-97074,
Germany }

\date{\today}

\begin{abstract}
We study two similar spin ladder systems with the ferromagnetic leg coupling.
First model includes two sorts of spins, $s= 1/2$ and $s= 1$, and  the second
model comprises only $s=1/2$ legs coupled by a "triangular" rung exchange.
We show that the antiferromagnetic (AF) rung coupling destroys the long-range 
order and eventually makes the systems equivalent to the AF $s=1/2$ Heisenberg 
chain. We study the crossover from the weak to strong coupling regime 
by different methods, particularly by comparing the results of the 
spin-wave theory and the bosonization approach.  We argue that the crossover 
regime is characterized by the gapless spectrum and non-universal critical 
exponents are different from those in XXZ model. 
\end{abstract}


\maketitle

\section{introduction}

The strongly correlated systems in one spatial dimension (1D)
attracted
an enormous theoretical and experimental interest last decade.
The 1D fermionic and spin systems were recognized long ago as the useful
theoretical models, where the interaction effects are very important and at
the same time are subject to rigorous analysis. \cite{GoNeTs,SchulzCuPi}
The experimental discovery
of the systems of predominantly 1D character inspired the renewed interest
to this class of problems. Among the experimental realizations
of the 1D systems one can mention the Bechgaards salts,
carbon nanotubes, copper oxides spin ladders and purely organic spin chain
compounds. \cite{GoNeTs}

While the physics of purely one-dimensional objects, or chains,
is well understood, \cite{GoNeTs,SchulzCuPi}
the spin ladders are still under intense investigation.
Even the isolated spin chains reveal a variety of unusual phenomena,
including the Haldane gap, spin-Peierls transition, magnetization
plateaus.
The ladders, consisting of a few coupled spin chains are generally
much richer in their behavior, and pose additional theoretical problems.

The interest to the problem of ladders may be traced back
to the earlier attempts to construct the continous representation
of spin $S=1$ variable in 1D out of $s=1/2$ quantities.
\cite{Schulz86}
The methods elaborated in these studies are now widely used in the
analysis of the ladder systems.

The basic model for the spin chain, the antiferromagnetic (AF) Heisenberg
$s=1/2$ chain, is thoroughly studied by various methods.
\cite{GoNeTs}
This may be one of the reasons, why a majority of
the theoretical papers discussing the spin ladders are now
confined to the treatment of quantum spin $s=1/2$ with antiferromagnetic
interaction along the legs.
Spin ladders with different spins or with a
ferromagnetic leg exchange attracted much less attention.

The spin chains and ladders consisting of different spins
\cite{organic} and with the
AF leg exchange were considered recently in
\cite{Ivanov01,Brehmer97,Wu99,TruGa01,Sakai02}.
Particularly,
the ferrimagnetic chains with alternating spin-1/2 and spin-1 were discussed
there. It was shown that, contrary to the case of equal spins,
the uncompensated spin value in a unit cell leads to the
gap in the spectrum and the appearance of the long range magnetic order.
\cite{Tian97}

The spin-1/2 ladder with ferromagnetic exchange along the legs was studied
in \cite{Roji96,Vekua03}. A rather rich phase diagram was demonstrated,
depending on
the details of the magnetic anisotropies for the leg and rung couplings.

In the present paper we study the mixed spin ($S=1$, $s=1/2$) spin ladder with
the ferromagnetic exchange along the legs and antiferromagnetic interaction on 
the rungs.In the absence of the rung coupling,
the individual chains show the ferromagnetic long-range order (LRO),
and the ground state is classical.
The spectrum and the ground state energy
is well described in terms of the linear
spin-wave theory. \cite{Ivanov01} We show
that the inclusion of the antiferromagnetic rung
coupling drives the system into the quantum regime, understood in
terms of the
AF Heisenberg spin-1/2 chain. In this regime the magnetic LRO is absent,
and the spatial correlations show the power-law decay.
Note that the uncompensated spin in a unit cell leads to the absence
of the gap in low-lying excitation spectra.

This primary observation is interesting on its own, because two limiting cases 
of the model allow the asymptotically exact solutions with gapless spectra. 
Hence our further motivation to study the crossover region starting from both 
limiting points. 

We discuss the regimes of weak and strong rung coupling for different types
of the leg exchange anisotropy. The consideration is somewhat complicated
by the absence of the established routines for our case.
The bosonization, an extremely useful tool in dealing with AF s=1/2 systems,
does not fully work here on two reasons. First is the existence of two sorts
of spins in a unit cell, and another is the ferromagnetic leg exchange.

Hence we supplement our study by
the consideration of a similar model, written
entirely for $s=1/2$ but with the modified rung couplings.
Using these models and comparing different approaches,
the spin-wave theory and bosonization,
we arrive at the unified description of the ferrimagnetic spin
ladders.
Particuarly, we discuss the spectrum and the correlations and
observe the crossover from the
weak to strong coupling regime. An attention is paid to a
subtler point in the bosonization procedure, a seemingly unstable Gaussian
effective action near the ferromagnetic point.

The rest of the paper is organized as follows. We discuss the mixed spin model
in strong and weak rung coupling regime in Sec.\ \ref{sec:mixed}.
The spin $s=1/2$ ladder with "triangular" rung exchange is introduced and
analyzed in Sec.\ \ref{sec:tria}. The existence of different
order parameters in a system is discussed in Sec.\ \ref{sec:OP}.
The discussion and conclusions are in Sec.\ \ref{sec:conc}

\section{ mixed spin ladder}
\label{sec:mixed}

We investigate the properties of a ladder system, consisting of
two sorts of spins, $s=1/2$ and $S=1$, arranged in a checkerboard
manner. The unit cell comprises four spins and the Hamiltonian is

        \begin{eqnarray}
        {\cal H} &=& - \sum_{i}J_\| ^\alpha
        \left (
        {s}_{1,2i}^\alpha { S}_{1,2i\pm1}^\alpha
        + { s}_{2,2i\pm1} ^\alpha { S}_{2,2i}^\alpha \right )
        \nonumber \\ &&
        + J_\perp \sum_i \left(
        {\bf s}_{1,2i}{\bf S}_{2,2i}
        +{\bf S}_{1,2i+1}{\bf s}_{2,2i+1} \right)
        \label{Ham1}
        \end{eqnarray}
with the first subscript (e.g.\  1 in ${\bf s}_{1,2i}$) labeling the leg, and
the second one denoting the site on it (odd or even).
Below we mostly consider the case of the AF rung coupling,
$J_\perp >0$.  The overall ferromagnetic exchange along the chains allows the
uniaxial anisotropy, $J_\|^x = J_\|^y =  J >0$,  $J_\|^z =J + D > 0$, $|D| \ll
J$.  In what follows, we consider also a useful generalization to higher spins,
$s \gg 1$ with the difference  $S-s=1/2$ kept fixed. The whole consideration is
done for zero temperature.

First let us briefly describe two simple limiting cases.
 At $D=0+$ and  $J_\perp =0$, we have two
ferromagnetic chains, which possess the long-range magnetic order. The spectrum
is quadratic at small wave vectors, $\omega \sim Jq^2$. We put the lattice
spacing $a$ to unity everywhere except the Sec.\ \ref{sec:tria-corr}.

\begin{figure}
\includegraphics[width=8cm]{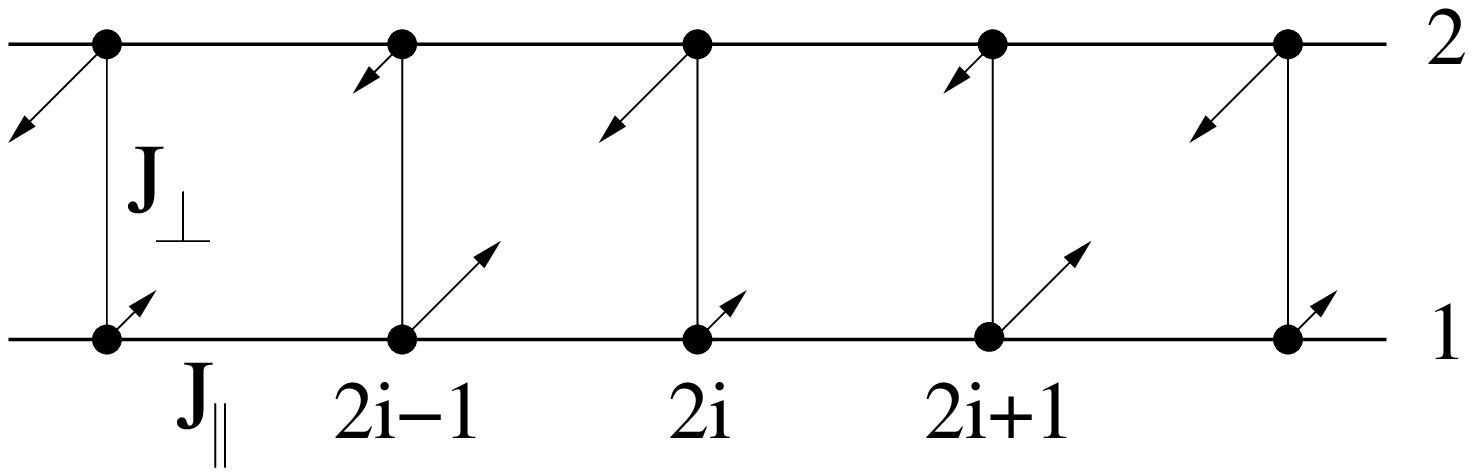}%
\caption{A ladder of spins $S=1$, $s=1/2$ \label{Fig:mixed}}
\end{figure}

In the opposite limiting case, $J_\perp \gg J $, the ground
state of two spins on the rung, say, ${\bf s}_{1,2i}$ and ${\bf
S}_{2,2i}$ is doublet, described by a  spin $s=1/2$ variable
${\bf \sigma}_{2i}$. We show below, that the effective
interaction between these ${\bf \sigma}_{i}$ is antiferromagnetic
for the above choice $J > 0$ and hence the situation is mapped
onto a well-known problem of $s=1/2$ Heisenberg antiferromagnet.
One does not expect the long-range order at $D=0$ and the spectrum
is linear $\varepsilon_k \sim  J k$. The correlation functions for this
case are described below.

The intermediate situation, $J \sim J_\perp $, $D\neq0$, is harder
to analyze and we present several ways to discuss it below.

\subsection{Strong rung coupling}

Consider first the case of strong perpendicular coupling,
$J_\perp \gg J $. Taking first $J=D=0$, one sees that each pair
of spins $s, S$ is characterized by a total spin ${\bf j} = {\bf
s} +{\bf S}$ and by the rung energy $E_j = J_\perp  {\bf s}{\bf S} = \frac
12 J_\perp (j(j+1) - s(s+1)-S(S+1)) $. The lowest state here is
doublet, $j=1/2$, the  first excited state is quadruplet $j=3/2$,
with  $ E_{3/2}-E_{1/2} =  3J_\perp/2$. The wave function of a
multiplet is

        \begin{equation}
        | j,m\rangle =\sum_{m_1,m_2} C^{jm}_{sm_1Sm_2}
        |sm_1\rangle |S m_2\rangle
        \label{klebsh}
        \end{equation}
with Clebsch-Gordan coefficients $C^{jm}_{sm_1Sm_2}$. The
interaction along the chain is considered now as a perturbation.
The formula (\ref{klebsh}) and consideration below are applicable for larger
spins as well.
In this more general case of  $s=S-1/2\geq 1/2$, the operators $s^\alpha$,
$S^\alpha$ act within the lowest doublet and connect the doublet with the
quadruplet, but the direct transitions to the higher states, $j=1/2 \to j\geq
5/2$, are absent. One can check that the corresponding matrix elements
are given by

        \begin{eqnarray}
        \langle j=1/2, m | s^\alpha |j=1/2,m'\rangle &=&
        -\frac{s}3 \sigma^\alpha _{mm'},
        \nonumber \\
        \langle j=1/2, m | S^\alpha |j=1/2,m'\rangle &=&
        \frac{S+1}3 \sigma^\alpha _{mm'},
        \label{matel}
        \end{eqnarray}
with $\sigma^\alpha _{mm'}$ the Pauli matrices. For $s=1/2$ this reads
        \begin{equation}
         s^\alpha \leftrightarrow -\frac16  \sigma^\alpha ,
        \quad
         S^\alpha \leftrightarrow \frac23
         \sigma^\alpha .
         \label{effSpins}
        \end{equation}

It shows that if
we consider only the projection of the spin operators onto the
lowest doublet, then the above Hamiltonian corresponds to the AF
Heisenberg spin-$1/2$ model of the form :
        \begin{equation}
        {\cal H}_{eff} = \sum_i J_{eff}^\alpha
         \sigma_i^\alpha  \sigma_{i+1}^\alpha,
         \label{Heff}
        \end{equation}
where
        \begin{equation}
        J_{eff}^\alpha (J_\perp \to \infty)=  2 J_\|^\alpha  s(S+1)/9.
        \label{Jeff0}
        \end{equation}
Below we will refer to this estimate of $J_{eff}^\alpha$
as $J_{eff}^{\alpha}(\infty)$.
For $s=1/2$ we have $J_{eff} ^\alpha (\infty)=2J_\|^\alpha /9$.

Let us next consider
the role of the higher states on a rung.
 We do not list here the matrix elements of the spin
operators $S^\alpha$, $s^\alpha$ for the transitions $j=1/2 \to
j=3/2$. We only note that they are proportional to
$\sqrt{s(S+1)}$ and are the same for $S^\alpha$ and $s^\alpha$, except for
the sign. The second-order correction in $J_\|$ between the adjacent rungs,
labeled by  1 and 2 below, can
be written as
        \begin{equation}
        J_\|^\alpha J_\|^\beta
        \frac{4
        \langle \frac12| s_1^\alpha | \frac32\rangle
        \langle \frac32| s_1^\beta | \frac12\rangle
        \langle \frac12| S_2^\alpha | \frac32\rangle
        \langle \frac32| S_2^\beta | \frac12\rangle
        }{2E_{1/2}-2E_{3/2}},
        \end{equation}
with factor 4 coming from the above property of coincidence of
the matrix elements. Noting that
       \begin{eqnarray}
        &&
       \langle \frac12,m| s_1^\alpha |\frac 32\rangle
       \langle \frac32| s_1^\beta |\frac12,m'\rangle
       \nonumber \\ &&=
        \frac{s(S+1)}{9} (
       2\delta_{\alpha\beta} \delta_{mm'} -i\epsilon_{\alpha\beta\gamma}
       \sigma^\gamma_{mm'} ),
       \label{transi}
       \end{eqnarray}
we find that the second order in $J_\|$ results in
(\ref{Heff}) with the renormalized value of effective interaction :
         \begin{equation}
         J_{eff}^\alpha (J_\perp)
         =  J_{eff}^{\alpha}(\infty)
         + \epsilon^2_{ \alpha\beta\gamma}
         \frac {J_{eff}^{\beta}(\infty) J_{eff}^{\gamma}(\infty) }
         {9J_\perp}.
         \label{Jeff}
         \end{equation}
with $J_{eff}^{\alpha}$ from (\ref{Jeff0}).
For the above parameters $J_\|^\alpha$ and $s=1/2$ one has
        \begin{eqnarray}
         J_{eff}^x &=& J_{eff}^y = \frac29 J
         + \frac {8J(J+D)}{729 J_\perp}, \\
         J_{eff}^z &=&  \frac29 (J+D)
         + \frac{8J^2}{729 J_\perp},
        \end{eqnarray}
which particularly means that the relative value of anisotropy decreases with
decreasing $J_\perp$.
The perturbation theory is expected to break down when the correction in
$J_\perp^{-1}$ is comparable to the first term in (\ref{Jeff}), which
happens roughly at  $J_\perp \sim  J_\|^\alpha s^2$. Note that this criterion
also corresponds to the point where the
bandwidth induced by $J_{eff}$ becomes comparable to the separation
$\sim J_\perp$ between the quadrupet and doublet.
At larger $J_\perp$ the low-energy dynamics is described  by the AF
spin-one-half XXZ model (\ref{Heff}) which is exactly solvable.

In the isotropic case,  $D=0$, the spectrum is linear,
$\omega =  \frac{\pi}{8} J_{eff} |q|$
and  the correlations are of the form
         \[
         \langle
         \sigma^\alpha_i \sigma^\alpha_{i+n} \rangle \sim
         |n|^{-2} + (-1)^n |n|^{-1} ,
         \]
with the omitted minor logarithmic corrections. \cite{Aff-log}

Taking into account the matrix elements (\ref{matel}) we find that the
leading asymptotes in the isotropic,  $s=1/2, S=1$ case are :
         \begin{eqnarray}
         \langle s^\alpha_{1,i} s^\alpha_{1,i+2n} \rangle
         &\sim&  |2n|^{-1}
         \nonumber \\
         \langle s^\alpha_{1,i} S^\alpha_{1,i+2n+1} \rangle
         &\sim&  4 |2n+1|^{-1}
         \label{StrongCorr}
         \\
         \langle S^\alpha_{1,i} S^\alpha_{1,i+2n} \rangle
         &\sim&  16 |2n|^{-1}
         \nonumber
         \end{eqnarray}
Thus the long-range ferromagnetic order is absent, but the
correlations are ferromagnetic and slowly decaying. In addition, there is a
subleading sign-reversal asymptote and modulation depending
on the spin value ($s=1/2,1$).
The correlations between the spins in different chains are
slowly decaying antiferromagnetic ones, e.g. $\langle s^\alpha_{1,i}
s^\alpha_{2,i+2n+1} \rangle \sim - |2n+1|^{-1} $. 
The above form of the correlations is obviously unchanged as long as $J_\perp 
\alt J$. 

\subsection{Weak rung coupling, spin-wave analysis}

Let us consider next the opposite
 limiting case, when the leg exchange dominates
and $J_\perp$ can be considered as perturbation.
First we explore the spin-wave formalism for the easy-axis
anisotropy, $D>0$. It was shown recently \cite{Ivanov01} that in case of
ferrimagnetic system in 1D the spin-wave description gives
very good estimate for the ground state energy and on-site
magnetization. Performing the standard Dyson-Maleyev expansion,
        \begin{eqnarray}
        S^z_1 &=& -S+a^\dagger  a , \quad  s^z_1 = -s+b^\dagger  b
        \nonumber \\
        S^+_1 &=& \sqrt{2S}
        a^\dagger  (1-a^\dagger a /(2S)) ,
        \nonumber \\
        s^+_1 & =& \sqrt{2s}b^\dagger  (1- b^\dagger b /(2s))
         \nonumber \\
       S^-_1 &=& \sqrt{2S}
        a , \quad
        s^- _1 = \sqrt{2s}b
         \nonumber \\
        S^z _2 &=& S- c^\dagger  c , \quad  s^z _2 = s- d^\dagger  d
        \nonumber \\
        S^+ _2 &=& \sqrt{2S}
          (1-c^\dagger c /(2S))c ,
          \nonumber \\
        s^+ _2  &=& \sqrt{2s} (1- d^\dagger d /(2s))d
         \nonumber \\
       S^- _2 &=& \sqrt{2S}
        c^\dagger , \quad
        s^- _2= \sqrt{2s}d^\dagger
        \end{eqnarray}
we write for the magnon Green function in the linear spin-wave theory (LSWT)
approximation
        \begin{eqnarray}
        \Phi^\dagger &=& (a_k^\dagger, b_k^\dagger, c_{-k}, d_{-k})
        \nonumber \\
        G_{ij}(k,t) &=& -i\theta(t) \langle
        [\Phi_i(k,t), \Phi^\dagger_j(k,0)] \rangle
        \end{eqnarray}
        \begin{widetext}
        \begin{eqnarray}
        G(k,\omega)^{-1} &=& -
        \begin{pmatrix}
        s(2J_\|^z +J_\perp)-\omega & \sqrt{sS} J\gamma_k
        & 0 &\sqrt{sS} J_\perp \\
        \sqrt{sS} J\gamma_k &S(2 J_\|^z +J_\perp)-\omega
        &\sqrt{sS}J_\perp&0\\
        0&\sqrt{sS} J_\perp& s(2 J_\|^z +J_\perp)+\omega&
        \sqrt{sS} J\gamma_k \\
         \sqrt{sS}J_\perp& 0 & \sqrt{sS} J\gamma_k
         &S(2 J_\|^z +J_\perp)+\omega
         \end{pmatrix}
     \label{HamLSWT1}
        \end{eqnarray}
\end{widetext}
Here $\gamma_k = 2\cos k$. The spectrum consists of doubly
degenerate acoustic and optical modes. The optical mode has the energy $\sim
\sqrt{sS} J $ for all wave vectors and the acoustic branch at small $k$ is
        \begin{equation}
        \varepsilon_{k} \simeq
        \frac {2s S}{s+S}\sqrt{
        (2D +J k^2) ( 2D +J k^2+2J_\perp )}
        \label{LSWT1}
        \end{equation}
so instead of the quadratic spectrum of purely FM case, we have
an approximately linear spectrum at small energies $\varepsilon_k \alt s
J_\perp $. The contribution of the zero-point fluctuations
into the average on-site magnetization can be estimated 
for $D\ll J_\perp \ll J$ as follows
        \begin{equation}
        s - \langle s_{1i} \rangle
        \sim \sqrt{J_\perp/J } \ln( \sqrt{J_\perp/D} ).
        \label{failureLSWT}
        \end{equation}
It means that the spin-wave approximation fails
when the latter quantity is of order of $s$. It happens roughly at
$J_\perp \agt s^2 J (\ln s^2  J/D )^{-2}$, which corresponds to
the crossover point $\gamma^\ast$ in \cite{Roji96}.
Apart from the logarithmic factor, this estimate agrees with the above
value $J_\perp
\sim s^2 J_\|$, obtained in the large $J_\perp$ limit. For lower $J_\perp $
the system shows the long-range order.

It might be instructive to consider the FM rung coupling $J_\perp < 0$. In this
case all the branches of the spectrum are gapful, the lowest modes are
          \begin{eqnarray}
        \varepsilon_{1,k} &\simeq&
        \frac {2s S}{s+S}
          ( 2D +J k^2),  \nonumber \\
         \varepsilon_{2,k} &\simeq&
        \frac {2s S}{s+S}
        ( 2D + 2|J_\perp| +J k^2 ) ,
        \label{disp14}
        \end{eqnarray}

Note that at the isotropic point $D=0$,  the long-range FM order and hence the
applicability of the LSWT is lost for any  $J_\perp>0$. The spectrum
(\ref{LSWT1}) in this case is gapless and linear with the spin-wave velocity
$\sim s\sqrt{J J_\perp}$, the limit of $s\simeq S\gg 1$ is assumed. We know that
increasing $J_\perp$ we should eventually recover the effective model
(\ref{Heff}), characterized by spinon velocity $\sim s^2 J$. These two
velocities match again at $J_\perp \sim s^2 J$.

Concluding this section, we also present the LSWT results for the lowest
branches of dispersion for the case of easy-plane anisotropy, $D<0$.
For the AF sign of the exchange $J_\perp$ one has
        \begin{eqnarray}
        \varepsilon_{1,k} &\simeq&
        \frac {2s S}{s+S}\sqrt{
         J k^2  ( 2|D| +J k^2+2J_\perp )},  \nonumber \\
         \varepsilon_{2,k} &\simeq&
        \frac {2s S}{s+S}\sqrt{
        ( 2|D| +J k^2 ) ( 2J_\perp +J k^2 )} ,
        \label{disp12}
        \end{eqnarray}
while for the FM exchange $J_\perp <0$ we obtain
        \begin{eqnarray}
        \varepsilon_{1,k} &\simeq&
        \frac {2s S}{s+S}\sqrt{
         J k^2  ( 2|D| +J k^2 )},
         \label{disp13}
         \\
         \varepsilon_{2,k} &\simeq&
        \frac {2s S}{s+S}\sqrt{
        ( 2|J_\perp|+J k^2 ) (2|D| + 2|J_\perp| +J k^2 )} ,
        \nonumber
        \end{eqnarray}
In this case LSWT is formally inapplicable, but the above formulas might be
useful for a comparison with further results.

Summarizing, we observe that while the
LSWT cannot treat the correlations correctly and formally is inapplicable in
the absence of the LRO, it provides simple and reasonable formulas for
the excitation spectra in a complicated one-dimensional ladder. We
illustrate this point below by discussing the spin-1/2 ladder, where
a rigorous description of the low-energy action is available.

\subsection{Equations of motion}
 
This subsection is devoted to the macroscopic equations describing the 
behavior of spins in two coupled chains. To make our calculations more 
transparent, we define new operators on a rung as superpositions of two spin 
operators        

        \begin{eqnarray}
        Q^\alpha_{2i}&=&S^\alpha_{2,2i}+s^\alpha_{1,2i},\quad
        R^\alpha_{2i}=S^\alpha_{2,2i}-s^\alpha_{1,2i}, \\ 
        Q^\alpha_{2i+1}&=&S^\alpha_{1,2i+1}+s^\alpha_{2,2i+1},\quad
        R^\alpha_{2i+1}=S^\alpha_{1,2i+1}-s^\alpha_{2,2i+1},
        \nonumber
        \end{eqnarray}
satisfying the following commutation relations:
       \begin{eqnarray}
       [Q_i^\alpha,Q_i^\beta] &=&
       i\epsilon_{\alpha\beta\gamma}Q_i^\gamma,\quad
       [R_i^\alpha,R_i^\beta]=i\epsilon_{\alpha\beta\gamma}Q_i^\gamma,
       \nonumber  \\ 
       {}[R_i^\alpha,Q_i^\beta] &=&i\epsilon_{\alpha\beta\gamma}R_i^\gamma.
       \label{ComRules}
       \end{eqnarray}
with $\epsilon_{\alpha\beta\gamma}$ totally antisymmetric tensor ;
the operators $Q^\alpha$ and $R^\alpha$ commute at different sites. 

The commutation relations (\ref{ComRules})
define the $SU(2)\bigotimes SU(2)=SO(4)$ group with two Casimir
operators given by
       \begin{equation} 
       {\bf Q\cdot R}=\frac{5}{4},\quad
       {\bf Q}^2+{\bf R}^2=\frac{11}{2}.
       \end{equation}
The Hamiltonian $H=H_\parallel+H_\perp$
takes the form
        \begin{eqnarray}
        \nonumber
        H_\parallel &=&-\frac{1}{2}
        \sum_i J_\parallel^\alpha\left(Q^\alpha_{i}Q^\alpha_{i+1}-
        R^\alpha_{i}R^\alpha_{i+1}\right)\\
        H_\perp&=&\frac{J_\perp}{4}
        \sum_i\left({\bf Q}^2_i-{\bf R}^2_i\right)=
        \frac{J_\perp}{2}\sum_i{\bf Q}^2_i +cst
        \end{eqnarray}
The equation of motions for  operators ${\bf Q}$ and ${\bf R}$
are given by
         \begin{equation}
         \dot{Q}^\alpha_j=i [H,Q^\alpha_j],\quad
         \dot{R}^\alpha_j=i [H,R^\alpha_j].
         \end{equation}
These equations correspond to the well-known Bloch
equations for precession of the magnetic moment
of ferromagnets (antiferromagnets). Taking into account that
the operators $Q^\alpha$ $(R^\alpha)$ are different on two 
different sublattices 
corresponding to odd(even) sites, we use the properties
          \[
          [{\bf Q}_i^2,Q_i^\alpha]= [{\bf R}_i^2,Q_i^\alpha]=0,
          \]
          \[
          [{\bf Q}_i^2,R_i^\alpha]= -[{\bf R}_i^2,R_i^\alpha]=
          i\epsilon_{\alpha\beta\gamma}\{R_i^\beta Q_i^\gamma\}
          \]
and adopt a symbolic notation
          \[
          i\epsilon_{\alpha\beta\gamma}\{R_i^\beta Q_i^\gamma\}=
          i\left({\bf R}_{i}\times{\bf Q}_{i}- 
          {\bf Q}_{i}\times{\bf R}_{i}\right)^\alpha .
          \]
Then we obtain the following system of coupled Bloch equations 
for the isotropic $J_\|^\alpha $
         \begin{eqnarray}
         \nonumber
         \partial_t{\bf Q}_{i} &=&
         \frac{J_\parallel}{2}\left( {\bf Q}_{i}\times
         {\bf Q}_{i\pm1}-{\bf R}_{i}\times{\bf R}_{i\pm1}\right)\\
         \nonumber
         \partial_t{\bf R}_{i} &=&
         \frac{J_\parallel}{2}\left({\bf Q}_{i}\times
         {\bf R}_{i\pm1}-{\bf R}_{i}\times{\bf Q}_{i\pm1}\right)+\\
         &+&
         \frac{J_\perp}{2}\left( {\bf Q}_{i}\times
         {\bf R}_{i}- {\bf R}_{i}\times{\bf Q}_{i}\right).
         \label{LL}
         \end{eqnarray}

Taking the continuum limit here, one has:
 \begin{eqnarray}
         \nonumber
         \partial_t{\bf Q} &=&
         \left( A_{\alpha\beta}{\bf Q}\times\frac{\partial^2
         {\bf Q}}{\partial x_\alpha\partial x_\beta}
         -B_{\alpha\beta}{\bf R}\times\frac{\partial^2 {\bf R}}
         {\partial x_\alpha\partial x_\beta}\right)\\
         \nonumber
         \partial_t{\bf R} &=&
         \left( B_{\alpha\beta} {\bf Q}\times\frac{\partial^2 {\bf R}}
         {\partial x_\alpha\partial x_\beta}-
         A_{\alpha\beta} {\bf R}\times\frac{\partial^2 {\bf Q}}
         {\partial x_\alpha\partial x_\beta}\right)+\\
         &+&
         C\left( {\bf Q}\times
         {\bf R}- {\bf R}\times{\bf Q}\right).
         \label{LL1}
         \end{eqnarray}
where for asymmetric two-leg ladder $\alpha=\beta=x$ and
$A_{xx}=B_{xx}=J_\parallel a^2/2,\;\; C=J_\perp/2$. The lattice index $i$ is 
omitted. Being re-written in terms of densities of magnetic moments, equations 
(\ref{LL1})correspond to the generalized Landau-Lifshitz equations for the 
dynamic $SO(4)$ group. Note that the operator ${\bf Q}$ represents a 
total spin on the rung, and the operator ${\bf R}$ does not have a simple 
interpretation.

We introduce the densities of magnetic moment characterizing two sublattices
         \begin{eqnarray}
         {\bf M}&=&\frac{2}{{\cal N}}\sum_{l}
         {\bf Q}_l,\quad
         {\bf N}=\frac{2}{{\cal N}}\sum_{l}
         (-1)^l{\bf Q}_l,
         \nonumber \\
         {\bf O}&=& \frac{2}{{\cal N}}\sum_{l}  
         {\bf R}_l,\quad
         {\bf T}= \frac{2}{{\cal N}}\sum_{l}  
         (-1)^l {\bf R}_l, 
         \end{eqnarray}   
with ${\cal N}$ being total number of rungs in the ladder.

Obviously, ${\bf M}$ and ${\bf N}$ represent the
uniform and staggered magnetizations along the ladder, whereas
${\bf T}$ and ${\bf O}$ can be interpreted as ``staggered rung''
magnetization in the uniform and staggered channels along the chain, 
respectively.
 The different ordered phases are 
charactererized by nontrivial values of
${\bf M,N,O,T}$ or some combinations of them whereas in disordered phase these
quantities are equal to zero. For example, the ordered FM phase is 
characterized by
${\bf M}\ne 0$. In Neel phase ${\bf N}\ne 0$. In ferrimagnetic phase both
${\bf M}\ne 0$ and ${\bf N}\ne0$. Therefore, the information about the order 
parameter
is necessary for deriving the macroscopic Landau-Lifshitz equations from 
the microscopic Bloch equations.

We point out that the scalar product of two spins on a rung
        $$
        {\bf s}_1{\bf S}_2=\frac{1}{4}\left({\bf Q}^2-{\bf R}^2\right)
        $$
may also be considered as a local order parameter (see discussion in 
the Section III).
        
Eqs.\  (\ref{LL}) are the central result of this subsection. 
Upon the assumption of the certain order parameter(s) they lead to macroscopic 
``equations of motion''. Applying a standard routine \cite{LL9}
one can show that these equations are equivalent to LSWT treatment 
and reproduce 
magnon dispersion laws discussed in the previous subsection. The detailed 
analysis of the Bloch equations for asymmetric $SO(4)$ ladders will be presented 
elsewhere. 

\section{triangular $s=1/2$ ladder}
\label{sec:tria}

\begin{figure}
\includegraphics[width=8cm]{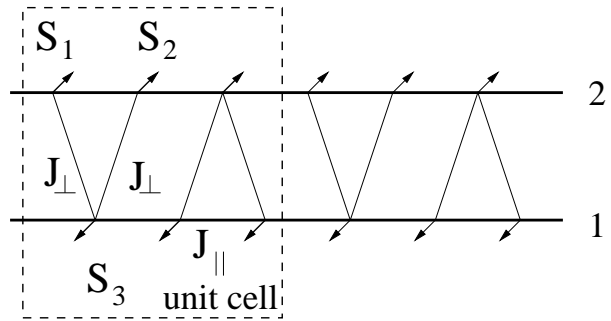}%
\caption{A triangular ladder of spins $s=1/2$ \label{Fig:tria}}
\end{figure}

One may regard spin 1 as a ground state triplet of two spins 1/2 coupled
ferromagnetically. Our model assumes that this triplet is also FM coupled to
other spins 1/2 on a chain.
Therefore instead of the FM chain of $n$ spins 1 and $n$
spins 1/2, one may consider $3n$ spins 1/2.  The model we propose is

        \begin{eqnarray}
        {\cal H} &=& - \sum_{i}J_\| ^\alpha
        \left (
        {s}_{1,i}^\alpha { s}_{1,i+1}^\alpha
        + { s}_{2,i } ^\alpha { s}_{2,i+1}^\alpha \right )
        \nonumber \\ &&
        + J_\perp \sum_i \left[
        {\bf s}_{1,3i} \left ( {\bf s}_{2,3i}+  {\bf s}_{2,3i+1} \right)
        \right.
    \nonumber \\ &&  \left.
        +\left( {\bf s}_{1,3i+1}+{\bf s}_{1,3i+2}\right)
        {\bf s}_{2,3i+2} \right]
        \label{triaHam}
        \end{eqnarray}
with the above choice of  $J_\|^\alpha$. The model
(\ref{triaHam}) is not equivalent to the previous one, eq.\ (\ref{Ham1}).
Indeed, the
exact mapping of (\ref{Ham1}) to (\ref{triaHam}) would include the strong
trimerized isotropic FM in-chain exchange, at those links which form the bases
of the triangles in Fig.\ \ref{Fig:tria}.  In that case one would first
consider the triangles and then couple them to each other, fully restoring the
consideration of the previous Section.
We show below that the model (\ref{triaHam}) with uniform value of the
rung exchange has two advantages.
First, it is equivalent to (\ref{Ham1}) at $J_\perp \to \infty$
and second, it is easier tractable in the opposite case $J_\perp \to 0$,
since the exact form of the low-energy action is available for the uniform
$J_\|^\alpha$.

\subsection{Strong rung coupling }

We consider a case when the AF exchange $J_\perp$ is much larger than
$J_\|^\alpha$, first for the isotropic $J_\|^\alpha=J_\|$.
In this case a main block is a triangle formed by two rungs, and the coupling of
triangles is a perturbation. The Hamiltonian for the triangle is
          \begin{equation}
          H_\triangle =
          J_\perp ({\bf s}_1 +{\bf s}_2) {\bf s}_3
          -J_\| ({\bf s}_1 {\bf s}_2 -1/4)
          \label{triangle}
          \end{equation}

The structure of the energy levels
is as follows. The term $J_\|$ groups the spins on one leg,
${\bf s}_1 $ and ${\bf s}_2$, into a triplet $|{\cal T}\rangle$ and a singlet
$|{\cal S}\rangle$. The singlet does not couple to ${\bf s}_3$ and results
in a total doublet denoted as $|D0\rangle$ with the energy $E_{D0} =J_\|$.
The triplet  $|{\cal T}\rangle$ of zero energy couples to ${\bf s}_3$
with the formation of doublet $|D1\rangle$ and quadruplet $|Q\rangle$.
The corresponding energies are $E_{D1} =  -J_\perp$ and
$E_{Q} = J_\perp/2$. For $J_\| \to \infty$ the state $|D0\rangle$
is unimportant, and ${\bf s}_1 $, ${\bf s}_2$ act as one spin $S=1$.
At the same time, the low-energy sector of the problem is associated with
the doublet $|D1\rangle$, and this doublet is the lowest state also for
the situation $J_\perp \gg J_\|$. Therefore the strong coupling limit of the
asymmetric ladder with two spins $s=1$, $s=1/2$ is described as well by
the triangular lattice depicted in Fig.(\ref{Fig:tria}) in the same limit. The
demand for in-triangle $J_\|$ to be large, in order to organize the effective
spin-1, is relaxed in this limit, and one can consider the situation  with the
uniform value of the exchange  $J_\|$ along the whole leg.

The presence of the anisotropy term $-D s^z_1 s^z_2$ in (\ref{triaHam})
is a negligible effect in the described picture. Indeed, this term translates
into a single-ion anisotropy of the triplet state, $(S^z)^2 $ and the
application of the formulas (\ref{matel}), (\ref{transi}) shows that it is only
the higher $|Q\rangle$  state, which
becomes split accordingly,  $\sim D (S^z)^2$.

Let us discuss the analog of Eq. (\ref{Jeff}) for the Hamiltonian
(\ref{triaHam}). In the limit $J_\perp \to \infty$ the effective interaction is
given by (\ref{Heff}) with $\sigma^\alpha$ acting within $|D1\rangle$ and
$J^{\alpha}_{eff}(\infty)
= \frac19 J^\alpha_\|$. The analog of (\ref{effSpins})
for eq. (\ref{triangle}) reads
        \begin{equation}
         s^\alpha_{1,2} \leftrightarrow \frac13 \sigma^\alpha,
        \quad  s^\alpha_3 \leftrightarrow -\frac16  \sigma^\alpha  .
         \label{effSpins3}
        \end{equation}

The second
order in $J^\alpha_\|$ corresponds to transitions to higher $|D0\rangle$ and
$|Q\rangle$ states. After some calculations we find
         \begin{eqnarray}
         J_{eff}^\alpha (J_\perp)&\simeq&  J_{eff}^{\alpha}(\infty)
                \label{Jeff2}  \\
         &&+  \epsilon^2_{ \alpha\beta\gamma}
         J_{eff}^{\beta}(\infty) J_{eff}^{\gamma}(\infty)
         \left[ \frac1{3J_\perp} -\frac3{5J_\perp + 2J } \right].
         \nonumber
         \end{eqnarray}
In (\ref{Jeff2}) the second term is negative for $J_\perp \agt J$, in contrast
to (\ref{Jeff}). It means that the value of $J_{eff}$ and the
relative exchange anisotropy $J_{eff}^z /J_{eff}^x -1$ increase
with decreasing of $J_\perp$.

Thus we conclude that the strong coupling limit of the the trinagular ladder
is described by the Hamiltonian (\ref{Heff})
with $J_{eff}$ given by (\ref{Jeff2}). Similarly to (\ref{StrongCorr}),
the in-leg correlations are slowly decaying ferromagnetic ones, although
their modulation is different due to eq.\ (\ref{effSpins3}) instead of
(\ref{effSpins}).

\subsection{Weak rung coupling, LSWT analysis}

For the case of ferromagnetic exchange with the easy-axis anisotropy we
employ the spin-wave formalism. The situation is complicated by the existence
of six spins in a unit cell. As a result, the quadratic
Hamiltonian is represented as $6\times6$ matrix. The Hamiltonian for the
interaction along the leg is standard, while the rung exchange needs some care.

Consider first two quantities $A_i$ and $B_i$ referring to $i$th site on the
upper and lower leg, respectively. If $A_i$ and $B_i$ are coupled by the
triangular rung exchange, Fig.\ \ref{Fig:tria}, then we have an expression
        \begin{equation}
        J_\perp
        \sum_{j} (A_{3j}+A_{3j+1}) B_{3j} + A_{3j+2} (B_{3j+1}+B_{3j+2}),
        \label{triarung}
        \end{equation}
with one term in the sum (\ref{triarung}) describing the coupling in the unit
cell, and three-site periodicity of the overall structure.
Going to Fourier components we have
        \begin{equation}
        \sum_q A_q (B_{-q} g_q +  B_{-q+\kappa} f_q +
        B_{-q-\kappa} f^\ast_{-q}),
        \label{triaRung}
        \end{equation}
with $\kappa  =2\pi/3$ and
        \begin{equation}
        g_q = \frac23 J_\perp(1+e^{iq} )\quad
        f_q = -\frac13 J_\perp(e^{i\kappa} + e^{iq-i\kappa} ).
        \end{equation}
Particularly, for $A_j = A = cst$, we obtain
        $$ \frac13 J_\perp
        A (4B_{0}  + B_{\kappa} +
        B_{-\kappa} ).
        $$

These preliminary notes show that  the rung interaction hybridizes
the magnons with the wave vectors $q$ and $q\pm \kappa$.
The LSWT Hamiltonian $H = \sum_k\Psi^\dagger_k H_k \Psi_k$
is obtained as a
matrix defined for the vector
\[ \Psi_k^\dagger = (a_{k-\kappa}^\dagger, a_{k}^\dagger,
a_{k+\kappa}^\dagger, b_{-k+\kappa}, b_{-k}, b_{-k-\kappa} ). \]
The Green function
      $ G_{ij}(k,t) = -i\theta(t) \langle
        [\Psi_i(k,t), \Psi^\dagger_j(k,0)] \rangle $
takes the form

\begin{widetext}
        \begin{equation}
        G(q,\omega)^{-1} = -
        \begin{pmatrix}
          \omega_{q-\kappa} +g_0+ \omega & f^\ast_\kappa& f_\kappa &
         g_{q-\kappa} & f^\ast_{\kappa-q} & f_{q-\kappa}\\
         f_\kappa& \omega_{q}+g_0 + \omega& f^\ast_\kappa& f_q& g_q&
         f^\ast_{-q} \\
          f^\ast_\kappa& f_\kappa& \omega_{q+\kappa}+g_0 + \omega &
         f^\ast_{-\kappa-q} &f_{q+\kappa} &g_{q+\kappa} \\
          g_{-q+\kappa} &f^\ast_{q} &f_{-q-\kappa}&
           \omega_{q-\kappa}+g_0 - \omega & f^\ast_0&  f_0\\
          f_{-q+\kappa} &g_{-q} & f^\ast_{\kappa+q} &f_0&
           \omega_{q} +g_0- \omega&  f^\ast_0\\
          f^\ast_{-\kappa+q} &f_{-q} &g_{-q-\kappa}&
         f^\ast_0&f_0 & \omega_{q+\kappa}+g_0 - \omega
         \end{pmatrix},
         \label{6x6}
        \end{equation}
\end{widetext}
where $\omega_{q} = J_\|^z - J_\|\cos q$ is the magnon spectrum
for isolated chains,
easy-axis anisotropy $J_\|^z-J_\|>0$ is assumed.
The new spectrum is determined from the
equation $det[G(q,\omega)^{-1}]=0$. The last equation amounts to the
third-order polynomial in $\omega^2$, which can be subsequently solved.

In order to analyze the lowest energies in the spectrum, $\omega\simeq0$ at
$q\simeq 0$ it is sufficient to deal with smaller matrices. It can be
shown that in this case one may consider almost degenerate
$2\times2$ block formed by second and fifth lines (columns). The asymptotic
expressions for the energies obtained this way coincide with those
obtained directly from (\ref{6x6}).

This simplified analysis can be also performed for other cases of the
in-chain exchange anisotropy and rung exchange. Particularly it is useful when
the analytic treatment of the spectrum becomes problematic. For instance, the
full LSWT consideration of the easy-plane $D<0$ for (\ref{triaHam}) amounts to
the analysis of $12\times12$ matrix Green's function, while the simplified
treatment reduces the calculation to bi-quadratic equation.

In the subsequent equations of this Section, the rung exchange $J_\perp$
appears with a prefactor $2g_{q=0}=8/3$. This prefactor is conveniently
incorporated into the quantity
    \begin{equation}
    J_1\equiv 8 J_\perp /3,
    \end{equation}
which is used below. Thus we interchangeably call $J_1$ and $J_\perp$
as the rung coupling value.

For the small anisotropy $D$, and $|J_1| \ll J$
we find the following asymptotic expressions. \\
i) Easy-axis, $D>0$, AF sign $J_\perp>0$. Doubly degenerate gapful mode.
        \begin{equation}
        \varepsilon_{1,2,k} \simeq
        \frac12 \sqrt{(2 D+Jk^2)(2 D+2 J_1 +Jk^2)}.
          \label{axisSWvelAF}
         \end{equation}
ii) Easy-plane, $D<0$, AF sign $J_\perp>0$. One gapless, one gapful mode.
        \begin{eqnarray}
        \varepsilon_{1,k}
        &\simeq &\frac12 \sqrt{Jk^2(2|D|+2J_1 + Jk^2)},
        \nonumber
        \\
        \varepsilon_{2,k} &\simeq &\frac12\sqrt{(2|D|+Jk^2)(2J_1 +Jk^2)}.
        \label{easySWvel}
        \end{eqnarray}
iii) Easy-axis, $D>0$, FM sign $J_\perp<0$. Two gapful modes.
        \begin{equation}
        \varepsilon_{1,k} \simeq D+\frac12 J k^2,
        \quad \varepsilon_{2,k} \simeq D+|J_1| +\frac12Jk^2.
        \label{axisSWvelFM}
        \end{equation}
iv) Easy-plane, $D<0$, FM sign $J_\perp<0$. One gapless, one gapful mode.
        \begin{eqnarray}
        \varepsilon_{1,k} &\simeq &\frac12\sqrt{Jk^2(2|D| + J k^2)},
        \label{easySWvelFM}
        \\
        \varepsilon_{2,k} &\simeq &
        \frac12\sqrt{(2|D|+2|J_1|+J k^2)(2|J_1| +J k^2)}.
        \nonumber
         \end{eqnarray}

\begin{figure}
\includegraphics[width=8cm]{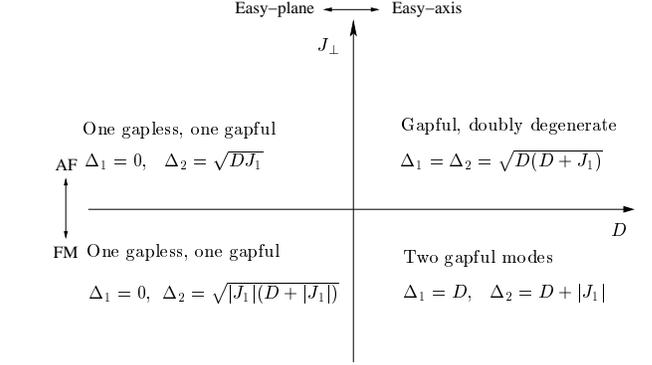}%
\caption{The character of dispersion in different domain of parameters,
the uniaxial anisotropy $D$ and the rung coupling $J_\perp$.
\label{Fig:diag}}
\end{figure}

These results are similar with eqs. (\ref{LSWT1}), (\ref{disp14}),
(\ref{disp12}), (\ref{disp13}) and will be compared below
with the treatment by bosonization. Particularly they show that the 
low-energy dynamics of the triangular ladder is similar to the mixed-spin ladder 
of Sec.\ \ref{sec:mixed} not only in the strong rung coupling regime, but also 
for the weak rung coupling. In Fig.\ \ref{Fig:diag} we depict the
character of dispersion in different domains of the small parameters, $D$,
$J_\perp$, according to eqs.\ (\ref{axisSWvelAF})-(\ref{easySWvelFM}).


\subsection{Weak rung coupling, bosonization}

As discussed above, the spin-wave
theory becomes inapplicable at larger $J_\perp$ when the role of quantum
fluctuations grows.
Instead, one may use the formalism which does not assume the average on-site
magnetization and is suitable for spin-one-half chains.
This formalism includes the Jordan-Wigner
transformation to spinless fermions, and the eventual continuum description with
the use of fermion-boson duality described elsewhere. \cite{GoNeTs}

This procedure is well defined for an easy-plane anisotropy, $D<0$
in (\ref{triaHam}), which is the case to be considered in this subsection;
the AF sign $J_\perp$ is implied. The
continuum representation of the spin operators in each of the chains
reads as
          \begin{eqnarray}
          s_x ^\pm&=& e^{\pm i\theta }(C + \cos(\pi x + 2\phi)),
          \label{spinRep}  \\
          s_x ^z&=& \pi^{-1} \partial_x\phi  + \cos(\pi x +2\phi)   ,
          \nonumber
          \end{eqnarray}
with the omitted normalization factors before cosines and constant
$C$ defined below. The bosonic fields $\phi_i(r)$, $\theta_i(x)$ are
characterized by the chain index $i$ and a continuous coordinate $x$.
The Hamiltonian $H= H_1 + H_2 +H_\perp$ has a part for the
noninteracting chains
         \begin{eqnarray}
          H_{i} &=&   \int dx\,
          \left(\frac {\pi u K}2 \Pi_i^2 +\frac{u}{2\pi K}
         (\partial_x \phi_i) ^2  \right)
         \label{onechain}
          \end{eqnarray}
where $i=1,2$ and
$\Pi_i = \pi^{-1}\partial _x \theta_i $ canonically conjugated momentum
to $\phi_i$. The form of $H_\perp$ in bosonization notation is discussed below.
The general form of the leg
Hamiltonian (\ref{onechain})
prescribed by the bosonization
procedure is complemented by the exact form of its coefficients, known from
Bethe Ansatz. \cite{LuPe75}
Denoting $\cos\pi\eta = J^z/J$, we have
     \[
      1/K=  2 \eta, \quad  u = J   (\sin\pi\eta) /(2 - 2\eta).
      \]
For  $|D| \ll J$ these formulas are simplified
      \begin{equation}
      \pi\eta \simeq \sqrt{\frac{2|D|}J},  \quad
      K\simeq \sqrt{\frac{\pi^2 J}{8|D|}} , \quad
      u \simeq \sqrt{\frac{|D|J}2 }.
      \label{bosoVel}
      \end{equation}
and the coefficient $C$ in (\ref{spinRep}) becomes \cite{LuZa} $C^2 \simeq
(\pi \eta)^{-\eta}/8\simeq 1/8$. Note that the spinon velocity $u$ in
(\ref{bosoVel}) coincides with the one obtained by LSWT, eq. (\ref{easySWvel})
at $J_1=0$. This unusual observation relates to the fact that the 
ferromagnetic ground state is describable in classical terms of the total 
magnetization, and the semiclassical LSWT approach should work well in the 
nearly ferromagnetic situation even without LRO. 

The rung interaction $H_\perp$ couples different terms of the spin densities.
However the interaction of the AF components, $\sim\cos(\pi x+ 2\phi)$ in
(\ref{spinRep}) is irrelevant in the renormalization group sense ; moreover,
the structure of  $H_\perp$, eq.(\ref{triaRung}), shows that this interaction
is absent in the lowest order.

It is convenient to introduce the symmetrized combinations  $\phi _\pm =
(\phi_1 \pm \phi_2 ) /\sqrt{2}$, $\theta _\pm = (\theta_1 \pm \theta_2 )
/\sqrt{2}$. In terms of these, the relevant and marginally relevant terms of
$H_\perp$ are
         \begin{eqnarray}
           H_\perp &\simeq &  J_1 \int dx\,
         [C^2\cos (\sqrt{2}\theta_- )
           + 8^{-1/2} C^2 \partial_x \theta_+ \sin (\sqrt{2}\theta_- )
          \nonumber \\
          && +
           (2\pi)^{-2} ((\partial_x \phi_+)^2 - ( \partial_x \phi_-)^2)
          ] .
          \label{bosoPerp}
          \end{eqnarray}
The second term in (\ref{bosoPerp}) comes from the gradient expansion of
(\ref{triaRung}). Its inclusion however does not change the Gaussian character
of the action for the field $\theta_+$ (see below). Integration over this latter
field produces a contribution $\sim (J_1^2/J) \cos 2\sqrt{2} \theta_-$, which
i) is less relevant and ii) has a smaller prefactor than the first term in
(\ref{bosoPerp}). That is why we omit the term $ \partial_x \theta_+ \sin
(\sqrt{2}\theta_-)$ below.

The remaining terms are combined into the Hamiltonian of the form
 $H = H_+ + H_-$ with
       \begin{eqnarray}
        H_{+} &=&   \int dx\,
       (\frac {\pi u_+ K_+}2 \Pi_+^2
        +\frac{u_+}{2\pi K_+}
        (\partial_x \phi_+) ^2  ),
        \label{bosHam2a} \\
         H_{-} &=&   \int dx\,
        (\frac {\pi u_- K_-}2 \Pi_-^2      +
        \frac{u_-}{2\pi K_-}
        (\partial_x \phi_-) ^2
         \nonumber \\ &&
         + J_1 C^2 \cos \sqrt{2}\theta_- ),
        \label{bosHam2b}
        \end{eqnarray}
where
         \begin{eqnarray}
         u_+ K_+  &=&  u_- K_- = u K \simeq \pi J/4,
         \nonumber \\
         u_\pm/  K_\pm &=& u / K  \pm J_1/2\pi \simeq
         (2|D|\pm J_1/2)/\pi.
         \label{newBosPar}
         \end{eqnarray}
Eqs. (\ref{bosHam2a}), (\ref{newBosPar}) show that the mode $\phi_+$ remains
gapless, and its velocity is increased with $J_\perp$, hence
it corresponds to $\varepsilon_1$ mode in (\ref{easySWvel}), (\ref{disp12}).

The situation with the $\phi_-$ mode is more complicated. Two features are noted
here, the instability of the Gaussian action at $J_1 > 4 |D|$ and
the appearance of the gap in the spectrum.

Indeed, the scaling dimension of the operator $\cos \sqrt{2}\theta_-$ in
(\ref{bosHam2b}) is $1/(2K_-) \ll 1$ and the dynamics of $\theta_-$ mode is
desribed by the sine-Gordon model in the quasiclassical limit, with a large
number of quantum bound states, or "breathers".
The gap $\Delta$ in the spectrum of $\theta_-$ field is
given by the mass of the lightest breather, which is roughly found  by expanding
the cosine term and rescaling the field   $\theta_- \to \theta_-/\sqrt{K_-}$
         \begin{equation}
         \Delta^2 \simeq 2\pi u_- J_1 C^2/K_-
         \simeq \frac{|J_1|}2  \left( |D| - \frac{J_1} 4\right)
         \end{equation}
In the leading order in $J_1$ this expresssion corresponds to
the mode $\varepsilon_2$ in (\ref{easySWvel}), (\ref{disp12}).
The refined value of the gap can be obtained after usual scaling
arguments \cite{AffOshi,Dmitriev02} or directly from the exact formulas
in\cite{LuZa}.
The identification of our model parameters with those of Lukyanov and
Zamolodchikov reads as
          \[
          \mu = u_- J_1 C^2 /2 , \quad \beta^2 = (4K_-)^{-1}
          \]
and $u_-$ stands for the overall energy scale.

The gap ($m$ in notation of \cite{LuZa}) is then found as
         \begin{equation}
         \Delta^2 \simeq
         \frac{J}2 \left( |D| - \frac {J_1}4 \right)
         \left ( \frac {|J_1|}{J} \right)
         ^{1/(1-1/(4K_-))}
         \label{gapCFT}
         \end{equation}
and the spectrum becomes
         \begin{eqnarray}
         \varepsilon_{+,k}^2&=& \frac 12
         \left(|D|+\frac{J_1}4\right) J k^2 ,\nonumber \\
         \varepsilon_{-,k}^2&=& 
         \frac 12
         \left(|D|-\frac{J_1}4\right)(|J_1|+ Jk^2),
         \label{bos-spec}
         \end{eqnarray}
The mean value of the cosine term is given by the expression
          \begin{equation}
          \langle \cos\sqrt{2}\theta_- \rangle \simeq
          -\left(\frac {\Delta}{4u_-}\right)
          ^{\frac{1}{2K_-}}\simeq
          -\left(\frac{|J_1|}{16 J }\right)^{\frac1{4K_-}}
          \label{averCos}
          \end{equation}

According to (\ref{newBosPar}), (\ref{averCos}), the increase of  $J_1$
leads to the instability of the Gaussian action, which happens
simultaneously with the saturation of the quantity  $\langle
\cos\sqrt{2}\theta_- \rangle$. Note that the similar situation was observed 
in \cite{Vekua03} for the simple two-leg ferromagnetic ladder.

For one chain, the breakdown of the Gaussian action happens at $D\geq 0$, and
corresponds to the transition to the ferromagnetic ground state. The average
value of spin in this case becomes $\langle s^z \rangle = \pi^{-1} \langle
\partial_x \phi \rangle = \pm 1/2$.

For a ladder, the discussed instability and the saturation of cosine term
correspond to the saturation of the scalar product of spins in different chains,
$\langle {\bf s}_{1,j} {\bf s}_{2,j} \rangle$ (see below). It means that
the spins in adjacent chains form a singlet state. The peculiarity of this
phenomenon is the energy scale when it happens,  $J_\perp \sim |D| $,
following from the bosonization (cf.\ \cite{Vekua03}). 

This small energy scale is unusual and may be compared to
the LSWT treatment.  Successful enough for isolated chains, LSWT
is in qualitative agreement with bosonization, regarding  the
increase of the spinon velocity with $J_\perp$ for the symmetric
mode $\phi_+$ as well as the gap value for the $\phi_-$ mode. At
the same time, LSWT predicts the {\em increase} in the velocity of
$\phi_-$  with $J_\perp$ at the energies higher than the gap
value. The bosonization says the opposite.

At this moment it is also instructive to consider the FM rung coupling $J_1 <0$.
In this case the LSWT formulas (\ref{disp13}), (\ref{easySWvelFM}) show again 
one gapless and one gapful mode, with the unchanged and increased velocities, 
respectively. The bosonization, (\ref{bos-spec}), provides the similar picture, 
but says again about the collapse of the gapless $\phi_+$ mode at $|J_1 |
\sim |D|$.

It is worth to note here that the average cosine term (\ref{averCos}) and 
the correlation length, eq.\ (\ref{corlength}) 
below, does not show any peculiarities at $J_1 \sim |D|$.  

A possible explanation for the above discrepancy stems from the
observation that $\varepsilon_2$ mode (\ref{easySWvel}) at $J_1 = |D|$ attains
the form  $\varepsilon_{2} \simeq |D| + Jk^2 /2 $. The region of linear
dispersion of bosons, a cornerstone of conformal treatment, is lost here,
which may be reflected by vanishing velocity in the bosonization
treatment. Note also, that eq.\ (\ref{easySWvel}) shows the roughly linear 
gapful spectrum upon the further increase of the rung exchange, $J_1 \gg |D|$. 
This feature should assumably be valid in the corrected bosonization treatment.

We suggest here that the action (\ref{bosHam2b}) should be complemented 
by the irrelevant terms, usually dropped in the infrared limit. They  
come from the consideration of the lattice Hamiltonian and are of the 
structure
         \begin{equation}
         a^2 [ (\partial_x \phi)^4 + (\partial_x^2 \phi)^2] ,
         \label{cubic}
         \end{equation} 
with $a$ 
the lattice spacing. The appearance of these terms is most easily 
observed by the consideration of one-chain XY model. In terms of the 
Jordan-Wigner fermions $\psi$ , one has the tight-binding fermionic 
spectrum, $\cos q$. Near the Fermi points $q=\pm\pi/2$ the leading terms 
in the expansion of the fermionic dispersion are the linear and cubic 
terms. The linear-in-$q$ fermionic term,$\sim \psi^\dagger \partial_x 
\psi$, transforms into $(\partial_x \phi)^2$ in the bosonic language, 
and the cubic term $\sim \psi^\dagger \partial_x^3 \psi$ attains the 
form (\ref{cubic}). Omitting the unknown coefficients $\sim a^2 \sim 1$
and denoting $\partial_x \phi_- = \phi_-'$ etc., 
the new Hamiltonian (\ref{bosHam2b}) is then schematically written as
  \begin{equation}
 \frac{J}2 \Pi_-^2 + \frac{J_1}2 \theta_-^2+ \frac{|D|-J_1}2
        (\phi_-')^2 +\frac{J}2 (\phi_-'')^2
        + \frac{J}4 ( \phi_-') ^4
       \label{scheme}
 \end{equation}
 
Let us consider first the case $J_1=0$. The interaction 
term $(\partial_x \phi)^4$ may be discarded in the 
infrared action, and the quadratic term $(\partial_x^2 \phi)^2$ modifies 
the spectrum, $\varepsilon_k^2 \sim |D|J k^2 + J^2 k^4$, so the spectrum 
may be regarded as linear only at $k \alt \sqrt{|D| /J}$. The latter 
estimate is in accordance with  the LSWT formulas(\ref{easySWvel}), 
(\ref{disp12}). The dynamical correlation function becomes 
\[ 
\langle \partial_x \phi_- ,\partial_x \phi_- \rangle_{k,\omega} \sim 
\frac{J k^2 }{\omega^2 - J|D| k^2 - J^2 k^4}
\]
which leads to the estimate for the average 
square (correlation function at $x=t=0$), 
$\langle (\partial_x \phi_- )^2 \rangle \sim  \int_0^1
dk\, k/\sqrt{k^2 + |D|/J} $. The latter quantity is defined by large $k 
\sim 1$ and shows that the fluctuations are strong, $\langle (\partial_x 
\phi_- )^2 \rangle \sim1 $, as should be expected from 
(\ref{spinRep}) for the fluctuating spins without LRO.

Consider now the case $J_1\neq 0$ in (\ref{scheme}).
In the regime with the negative coefficient before $(\partial_x \phi_-) 
^2$, the interaction term $(\partial_x \phi)^4$ stabilizes the action 
against the divergent static mean value $\partial_x \phi_-$. 
The usual recipe here is first to determine the variational static 
solution to the above Hamiltonian letting $\Pi_- = 0 = \theta_-$, see, 
e.g.\ \cite{solitons} and references therein. The trivial 
classical solution is the doubly degenerate vacuum $\partial_x 
\phi_-^{(0)} \equiv \rho_0 \sim \pm \sqrt{(J_1 -|D|)/J}$. 
The spectrum of fluctuations around it is well-defined with the velocity 
$u_\ast^2 \sim J (J_1 -|D|)$ and the Luttinger exponent $K_\ast^2 \sim J 
/(J_1 -|D|)$. The short-range fluctuations are still determined by the 
quadratic part of the spectrum, $\varepsilon_k \sim J k^2$, and the 
average square of the fluctuations is similarly estimated, $\langle 
(\partial_x \phi_- -\partial_x \phi_-^{(0)} )^2 \rangle \sim 1 $. It 
shows that the amplitude of the quantum fluctuations exceeds the 
distance between the vacua, $2\rho_0$, which makes the choice of the 
classical vacuum dubious. 

The refined analysis reveals the existence of multisoliton classical 
solutions to (\ref{scheme}). Variating the static Lagrangian 
over $\phi'(x)$ and letting $\phi'(x) = \rho_0 f(y)$ with 
$y=\rho_0 x$ we obtain an equation
   $   -{d^2 f}/{(dy)^2} = f -f^3 ,$
which allows a solution of the form $f = \alpha_1 {\rm sn}(\alpha_2 
y,\kappa) $ with ${\rm sn}(y,\kappa)$ the Jacobi elliptic function and 
$\kappa$ the elliptic index. 
It leads to the $N$-soliton solution for classical vacuum, 
 $\partial_x \phi_-^{(0)} = \pm \rho_0 
\sqrt{2\kappa^2/(1+\kappa^2)}{\rm sn}(x 
\rho_0/\sqrt{1+\kappa^2}, \kappa)$, with the soliton 
density $N/L = \rho_0 /(2 \sqrt{1+\kappa^2} K(\kappa))$. \cite{solitons}
In the limiting case $\kappa=1$ 
one has one soliton, $\partial_x \phi_-^{(0)} \sim \rho_0 
\tanh (x\rho_0/\sqrt{2})$. The difference in the classical energy 
between these $N$-soliton solutions and the above trivial vacua is 
estimated as $\sim N J \rho_0^{3}$, i.e. a small quantity at $N\sim 1\ll 
\rho_0 L $, as compared to the classical energy $\sim L J \rho_0^4$.
The full analysis 
of the problem should hence include the summation over the $N$-soliton 
solutions. The existence of the quantum gap, $\Delta^2 \sim  J_1(J_1 
-|D|)$, expected from the  $J_1 \theta_-^2$ term in (\ref{scheme}), only 
adds to the complexity of this problem, which should be dicussed 
elsewhere. One can only observe here that the necessity of summation 
over the classical vacua provides the absence of the staggered 
magnetization along the $z-$axis, associated with the non-zero classical 
$\partial_x \phi_-$. 

Knowing the spectrum and the Luttinger exponents, one can use the 
principal advantage of the bosonization in evaluation of the correlation 
functions. These correlations are discussed in the next section upon the 
assumption of the weaker coupling, $J_1 \alt |D|$.

\subsection{Correlation functions}
\label{sec:tria-corr}

The spectrum of (\ref{bosHam2b}) consists of one gapless and one gapful 
mode.The gap $\Delta$ corresponds to a finite correlation length
           \begin{equation}
           \xi = u/\Delta \sim \sqrt{J/J_1}
           \label{corlength}
           \end{equation}
separating domains of different behavior of the correlation functions.
The transverse spin correlations in one chain, $j=1,2$,  have the form
\cite{Luk97}
           \begin{eqnarray}
           \langle s^+_{j,0} s^-_{j,r} \rangle
          &\sim&
          r^{-1/4K_+}e^{ (K_0(r/\xi)-K_0(a/\xi))/( 4K_-)},
          \label{Xcorr11}
          \end{eqnarray}
with $K_0(x), K_1(x)$ modified Bessel functions and $a$ the lattice spacing.
At shorter distances, $r < \xi$, this expression becomes
  $\langle s^x s^x \rangle \sim r^{-1/ 4K_+ -1/ 4K_- }$ , while at large  $r
\agt \xi$ one has   $\langle s^x s^x \rangle \sim r^{-1/ 4K_+} \xi^{-1/ 4K_-}$.
The interchain correlations are
          \begin{eqnarray}
          \langle s^+_{1,0} s^-_{2,r} \rangle
          &\sim& -
          r^{-1/4K_+}e^{ - (K_0(r/\xi)+K_0(a/\xi))/( 4K_-) } ,
          \label{Xcorr12}
          \end{eqnarray}
with the behavior
$ -r^{-1/ 4K_+ + 1/ 4K_- }\xi^{-1/ 2K_-}$ and
$ -r^{-1/ 4K_+ }\xi^{-1/ 4K_-}$
at shorter and larger distances, respectively. Hence the interchain correlations 
decay faster beyond the scale $r\sim \xi$.

The longitudinal correlations are obtained in the form
           \begin{eqnarray}
          \langle s^z_{1,0} s^z_{1,r} \rangle
          &\sim&
          \frac{K_+} {r^{2}} + \frac{K_-}{\xi r } K_1(r/\xi) ,\\
            \langle s^z_{1,0} s^z_{2,r} \rangle
          &\sim&
          \frac{K_+} {r^{2}} - \frac{K_-}{\xi r } K_1(r/\xi)  ,
          \end{eqnarray}
which shows particularly that at $r <\xi$ the interchain correlations are of the
AF character.

The parameters $K_{+}^{-1}, u_+, \Delta, \xi^{-1}$  increase
with $J_\perp$. Hence the transverse correlations decay faster at
larger  $J_\perp$. We argued above that in the strong-coupling
limit $J_\perp \to \infty$ one deals approximately with the AF
Heisenberg chain situation, wherein $\langle s^x s^x \rangle \sim
r^{-1}$. Comparing it with (\ref{Xcorr11}), (\ref{Xcorr12}) one
may conclude that $K_+^{-1}$ should reach the value $1/4$ in the
strong coupling regime $J_\perp \sim J$. Actually it is not so
simple, since the derivation of (\ref{bosHam2b}) assumed $K_+
> 1$, and other terms of the rung interaction become important at
smaller $K_+$. As a result, one expects that the increase of
$J_\perp$ eventually changes the structure of the effective
low-energy action.

Summarizing, we show that the ``triangular'' model of 
this section is equivalent to one of Sec.\  \ref{sec:mixed} for the 
strong rung couplings. Further its dynamics is similar to one of the mixed spin 
ladder also for the weak rung coupling, as shown by LSWT approach 
complemented by the bosonization. Two latter techniques reveal certain 
shortcomings in the decription of the situation, as LSWT becomes formally  
inapplicable without LRO and the bosonization becomes unstable at the level of 
the Gaussian action.

Working in the close vicinity of the FM point in the parameter space 
$(D, J_\perp)$, we observe that the transition to the FM ordered phase is of the 
first order at the line $D\neq 0, J_\perp =0$. At non-zero AF values of 
$J_\perp$, this transition becomes the second-order one, at the line 
$J_\perp^\ast \sim D >0$. Approaching this transtion line from above, $J_\perp > 
J_\perp^\ast$, one should observe the divergence of the correlation length and 
vanishing critical exponents of the correlation functions. 
 
Combining the results of Sec.\  \ref{sec:mixed} and Sec.\ \ref{sec:tria}, we 
conjecture that the crossover from the weak to strong rung 
coupling regime for the isotropic situation, $D=0$, is characterized by the 
absence of the long-range order and gapless character of dispersion. Increasing 
the AF value of $J_\perp$ on has $\varepsilon_k \sim \sqrt{JJ_\perp} k$ until 
$J_\perp\alt J$ and $\varepsilon_k \sim J k$ at $J_\perp\agt J$. 
This form of dispersion takes place at $k \alt \xi^{-1} \sim \sqrt{J_\perp/J}$. 
The correlation functions are of the form $\langle s^\alpha_0 s^\alpha_r 
\rangle \sim r^{-\gamma}$ beyond the correlation length, $\xi$, with $\gamma 
\sim \sqrt{J_\perp/J}$ at $J_\perp \alt J$ and $\gamma =1$ otherwise. 

Note that this non-universal behavior of the critical exponent $\gamma$ 
characterizes the {\em isotropic} gapless situation and should be contrasted to 
the well-studied case of a gapless XXZ chain. \cite{GoNeTs} In the latter 
case one has different exponents $\gamma_\alpha$ for different spin projections 
$\alpha$, with certain relations between them, e.g. $\gamma_x \gamma_z = 1/4$. 

\section{Order parameters}
\label{sec:OP}

\subsection{String order parameter
vs.\ scalar product}

In the paper \cite{SNT96} (see also \cite{WEFN02,Nishiyama95}) a model 
of a symmetric AF Heisenberg ladder of spins $s=1/2$ was considered. 
Particularly, the authors discussed the string order parameter (OP), 
which was associated with the topological OP introduced earlier 
\cite{denNijsRom} for the spin-1 chain. In fact, the discussion by 
Shelton {\em et al.} \cite{SNT96} for non-zero AF rung exchange 
$J_\perp$ can be reduced to the observation that the scalar product 
${\bf s}_1 {\bf s}_2$ on the rung assumes the non-zero value.

Let us characterize
each state of two spins on a rung $j$ in terms of singlet
$|{\cal S}_j\rangle $ and triplet $|{\cal T}_j\rangle$. The ground state
$|G\rangle $ of the whole ladder has a component comprised of all rung
singlets, $|{\cal S}_{tot}\rangle  = \otimes _j |{\cal S}_j\rangle $. It is
clear that for the case of extremely large AF rung exchange the weight $W$ of
$|{\cal S}_{tot}\rangle$ in $|G\rangle $ is unity. One expects that for moderate
AF $J_\perp \sim J_\|$ this weight $W$ is finite. Consider now
the spin product on $j$th rung $-4 s^\alpha_{j,1}s^\alpha_{j,2}
= \exp i\pi (s^\alpha_{j,1}+s^\alpha_{j,2})$ with $\alpha =
x,y,z$, which may be represented as
       \begin{equation}
        -4 s^\alpha_{j,1}s^\alpha_{j,2} =
        {\cal P}_{s,j}+(1-{\cal P}_{s,j}) e^{i \pi S^\alpha_j}
        \label{rungProd}
       \end{equation}
with  ${\cal P}_{s,j}$ projecting onto the $j$th singlet and $S^\alpha_j$ spin-1
operator for the $j$th triplet. Note that the presence of  ${\cal P}_{s,j}$
makes (\ref{rungProd}) different from the operator $e^{i \pi S^\alpha_j}$ used
by den Nijs and Rommelse in their discussion \cite{denNijsRom}
of the spin-1 chain.

Indeed, the "string" operator $\prod _{j=l}^n (-4 s^\alpha_{j,1}s^\alpha_{j,2})$
has its ground-state expectation value contributed by the weight of the
$|{\cal S}_{tot}\rangle$ state. This partial contribution is equal to $W$ and
does not depend on the distance $(n-l)$. Particularly, the expectation value
of the scalar product $ -4{\bf s}_{j,1} {\bf s}_{j,2} = 4 {\cal P}_{s,j}-1 $
has a contribution $3W$ from  $|{\cal S}_{tot}\rangle$ state. In bosonization
notation we have, \[  \langle {\bf s}_{j,1} {\bf s}_{j,2}
\rangle \sim \langle\cos \sqrt{2}\theta_-\rangle   -  \langle\cos
\sqrt{8}\phi_+\rangle +  \langle\cos \sqrt{8}\phi_-\rangle . \]
For the AF signs of  $J_\|, J_\perp$, considered in \cite{SNT96},
first two cosines in the latter expression have nonzero values.
Some inspection shows that these values correspond to ones reported in
\cite{SNT96} for the infinitely long string OP.

Hence we conclude that the string OP discussed in
\cite{SNT96,WEFN02} for AF rung interaction
can be identified with the scalar product of spins on a rung
and measures the weight $W$ of the total singlet in the ground
state. It should be stressed, that our above arguments are not 
applicable for the FM rung interaction, when the lowest rung state is 
triplet. In this latter case the string order parameter discussed in 
\cite{SNT96,WEFN02,Nishiyama95} is the appropriate description and 
cannot be reduced to a local scalar product.

Clearly, the non-zero average value of the scalar product is disconnected from
the appearance of the on-site magnetization, as discussed below.

\subsection{Asymmetric ladders}

Applying the same type of consideration to our above systems, we can say, e.g.,
that for the mixed spin ladder the order parameter is the average value
of the scalar product on the rung $p_j = {\bf S}_j {\bf s}_j$. It assumes two
values, $p_j = -1$ and $1/2$ for the rung doublet and quadruplet, respectively.
For $J_\perp = 0$ all these six states have the same weight, resulting in
$ \langle p_j \rangle= 0$. With the increase of AF rung exchange, $\langle p_j
\rangle$ saturates into $-1$ value.

Similarly, for the triangular ladder, one considers the combined scalar
product $p_\triangle = ({\bf s}_1 +{\bf s}_2) {\bf s}_3 $, see
Eq.(\ref{triangle}). This quantity takes three possible values $p_\triangle =
-1, 0, 1/2$ in the states $|D1\rangle, |D0\rangle , |Q\rangle$, respectively.
Increasing $J_\perp$, the $|D1\rangle$ state becomes favorable, with
$\langle p_\triangle \rangle \to -1 $.

Notice,  that the discussed order parameter is bilinear
in spins, independent of the in-leg spin exchange anisotropy and does not imply
the ordering of individual spins.

The spin ordering in a proper sense depends
on the sign of the uniaxial anisotropy. Particularly, in the case of the
easy-plane anisotropy, both the weak and strong rung coupling regimes correspond
to XXZ model in the absence of LRO. Therefore one does not expect the spin
ordering here.

The case of the easy-axis anisotropy can be analyzed for the mixed
spin model.  We showed in Sec.\ \ref{sec:mixed} that the LSWT, applicable for
isolated chains, fails for the intermediate $J_\perp$. At the same time, the
strong coupling Hamiltonian (\ref{Heff}) is the AF easy-axis XXZ model.
It means the appearance of non-zero
staggered magnetization for the effective spins  $\sigma^z_j$ in
(\ref{Heff}). Scaling estimates (see e.g. \cite{Dmitriev02}) show that $\langle
\sigma^z_j \rangle \sim (-1)^j (D/J)^\alpha$ with $\alpha = (\pi/4)\sqrt{J/D}$.
This exponentially small value of the order parameter for the effective
Hamiltonian (\ref{Heff})
translates into the corresponding values for initial spins according
to (\ref{effSpins}). Note that the average spins in one leg are aligned in one
direction, but due to the difference in their contribution to the rung doublet
state, eqs.\ (\ref{effSpins}), (\ref{effSpins3}), both the uniform and the
staggered magnetization is present in the system. \cite{Tian97}

\section{conclusions}
\label{sec:conc}

We demonstrate above that the mixed spin ladders and triangular ladders
with the ferromagnetic coupling along the legs are generic models for 
description of a transition from the classical (ferrimagnetic) to quantum 
(antiferromagnetic) regime. The individual legs with the isotropic Heisenberg 
exchange show the classical groundstate and their dynamics is well described by 
the quasiclassical spin-wave theory. Turning on the AF rung coupling introduces 
strong fluctuations, which destroy the long-range order and eventually make the 
system equivalent to the quantum AF spin $s=1/2$ Heisenberg model.

We showed that in a large domain of parameters for these ladders
the spin wave theory, although missing certain features caused 
by quantum fluctuations in one dimension,   is still
quite instructive for the qualitative determination of the spectra,
which allows the further comparison with more sophisticated
methods. The refined analysis of the spectrum and correlations by the
bosonization technique complements the investigation
of the "quantum" regions of the phase diagram. As a result, the unified
description of the model becomes possible, partly including the
complicated crossover region from the weak to strong rung coupling limit.

We argue that for the isotropic spin exchange this crossover
is characterized by the gapless spectrum with (spinon) velocity $\sim 
\sqrt{JJ_\perp}$. The vanishing velocity at $J_\perp =0$ corresponds to the 
first-order phase transition to the ferromagnetic state. The asymptotic 
decay of correlation functions is described by unique critical exponent, 
$\gamma \sim \sqrt{J_\perp/J}$, for all three projections of spin. 
This type of behavior makes the mixed spin ladder in its crossover regime quite 
distinct from the AF  $s=1/2$ Heisenberg model, which would be very interesting 
to verify by independent, e.g. numerical, methods.

\begin{acknowledgments}

We thank A.Katanin, K.A.Kikoin, A.Luther, A.Muramatsu for useful discussions.
This work is partially supported by SFB-410 project and the
Transnational Access
program $\#$ RITA-CT-2003-506095 at Weizmann Institute of Sciences (MNK).

\end{acknowledgments}


\end{document}